\documentstyle[onecolumn]{mn}
\input{psfig.sty} 
\newcommand{\reference}{\bibitem}
\def\beq{\begin{equation}}
\def\eeq{\end{equation}}
\def\bey{\begin{eqnarray}}
\def\eey{\end{eqnarray}}
\def\beqarray{\begin{eqnarray}}
\def\eeqarray{\end{eqnarray}}

\def\mpc{\,{\rm {Mpc}}}

\def\mpch{\,h^{-1}{\rm {Mpc}}}
\def\kms{\,{\rm {km\, s^{-1}}}}
\def\msun{\,\rm M_\odot}
\def\Kdegree{{\,\rm K}}

\def\v200{V_{200}}

\def\LCDM{\rm \Lambda CDM}

\def\omnow{\Omega_{\rm m, 0}}

\def\ovnow{\Omega_{\rm \Lambda,0}}

\def\rmd{{\rm d}}

\input epsf
\title[]        
{The abundance and clustering of dark haloes in 
the standard $\Lambda$CDM cosmogony}
\author[Mo \& White]
{
H. J. Mo\thanks{e-mail: hom@mpa-garching.mpg.de}
and S.D.M. White\thanks{e-mail: swhite@mpa-garching.mpg.de} \\
Max-Planck-Institute f\"ur Astrophysik,
        Karl-Schwarzschild-Strasse 1, 85740 Garching, Germany}
\date{Accepted ........
      Received .......;
      in original form .......}
\pubyear{2000}

\begin{document}
\maketitle
\begin{abstract}

Much evidence suggests that we live in a flat
Cold Dark Matter universe with a cosmological constant.
Accurate analytic formulae are now available for
many properties of the dark halo population in such a
Universe. Assuming current ``concordance'' values for
the cosmological parameters, we plot halo abundance against
redshift as a function of halo mass, of halo temperature, of
the fraction of cosmic matter in haloes, of halo
clustering strength, and of the clustering strength of the
$z=0$ descendants of high redshift haloes. These plots are useful for
understanding how nonlinear structure grows in the model.
They demonstrate a number of properties which may seem
surprising, for example: $10^9\msun$ haloes are as abundant at
$z=20$ as $L_*$ galaxies are today; $10^6$K haloes are equally
abundant at $z=8$ and at $z=0$; 10\% of all matter is currently
in haloes hotter than 1 keV, while more than half is in haloes 
too cool to trap photo-ionized gas; 1\% of all matter at
$z=15$ is in haloes hot enough to ionise hydrogen; haloes of
given mass or temperature are more clustered at
{\it higher} redshift; haloes with the abundance of present-day
$L_*$ galaxies are equally clustered at all $z<20$; the metals
produced by star-formation at $z>10$ are more
clustered at $z=0$ than are $L_*$ galaxies.
\end{abstract}


\section {INTRODUCTION}

Over the last 20 years the Cold Dark Matter (CDM) cosmogony
set out by Peebles (1982), Blumenthal et al. (1984) and
Davis et al. (1985) has 
become the standard model for structure formation in the Universe.
It assumes the cosmic mass budget to be dominated by
an as yet unidentified, weakly interacting massive particle, 
whose gravitational effects build 
structure from an initially Gaussian distribution of 
adiabatic fluctuations. All structure originated as quantum zero-point
fluctuations during an early period of inflationary expansion
(see Guth 1997 for a review).
CDM models are specified by a small set of parameters. These are
the fractions of the current critical density in CDM 
($\Omega_{\rm CDM,0}$), in baryons ($\Omega_{\rm B,0}$),
and in the cosmological constant ($\Omega_{\Lambda, 0}$), the
current expansion rate, specified by
Hubble's constant $H_0=100 h\kms\mpc^{-1}$, and the power
spectrum of the initial density fluctuations. Over the 
wavenumber range of interest, the latter can be approximated by a power-law
$P_i(k)=A k^n$ with $n$ near unity. 
Conventionally, the amplitude of the power spectrum is quoted 
in terms of $\sigma_8$, the {\it rms} (extrapolated) linear mass 
fluctuation at $z=0$ in a sphere of radius $8\mpch$.

These parameters are heavily constrained by observation
and a ``concordance'' model, the standard $\Lambda$CDM model, 
has now emerged, with
$\Omega_0\equiv \Omega_{\rm CDM, 0}+\Omega_{\rm B, 0}+\Omega_{\Lambda,
0} = 1$,
$\Omega_{\Lambda, 0} \sim 0.7$, $h\sim 0.7$,
$\Omega_{\rm B, 0}\sim 0.02 h^{-2}$, $n\sim 1$, and $\sigma_8\sim 0.9$.
The evidence that the Universe is flat ($\Omega_0= 1$) comes primarily
from recent measurements of the angular power spectrum of fluctuations
in the temperature of the Cosmic Microwave Background (de Bernardis et 
al. 2002). The non-zero value for the cosmological constant comes
from combining this result with the distance-luminosity relation of
type Ia supernovae (Perlmutter et al. 1999). 
The value of Hubble's constant is taken from the HST programme to measure
galaxy distances using Cepheids (Freedman et al. 2001).  The baryon 
density is derived by comparing cosmic nucleosynthesis calculations 
to the measured deuterium abundance in the intergalactic medium 
(O'Meara et al. 2001; Olive, Steigman \& Walker 2000
and references therein). Finally, the values of $n$ 
and $\sigma_8$ are based on 
COBE observations of the CMB temperature fluctuations on large
scales (Bennett et al. 1996). The predictions with this set 
of model parameters are consistent with a broad range of  other 
observations, most notably with the number density of rich clusters
of galaxies (White, Efstathiou \& Frenk, 1993; 
Viana \& Liddle 1999; Reiprich \& B\"orhinger 2002), 
with the shear field produced by weak
gravitational lensing (e.g. Van Waerbeke et al. 2001), with
the clustering of galaxies and clusters on large scales
(Mo, Jing \& White 1996; Jing, Mo \& Boerner 1998; 
Benson et al. 2000; Peacock et al. 2001; 
Schuecker et al. 2001; Efstathiou et al. 2002), and with 
structure in the high redshift intergalactic
medium as measured using Ly$\alpha$ forest absorption 
in quasar spectra (Croft et al. 1999). 
We note that full consensus
on a ``concordance'' model is still lacking. 
For example, most determinations of Hubble's constant from time-delay
measurements in gravitational lensed quasars give values of $h$ lower 
than $0.7$ (Kochanek 2002, but compare Hjorth et al 2002); some 
recent estimates of $\sigma_8$ from cluster abundance are
significantly lower than $0.9$ 
(e.g. Seljak 2001; Viana, Nichol \& Liddle 2002); and
the results of Croft et al. (1999) in fact favour 
$\Omega_{\rm CDM, 0}$ values rather higher than $0.3$.  

In the CDM cosmogony, a key concept in the build-up of structure
is the formation of dark matter haloes. These are quasi-equilibrium 
systems of dark matter particles, formed through non-linear gravitational
collapse. In hierarchical scenarios like CDM, most mass 
at any given time is bound within dark haloes; galaxies and 
other luminous objects are assumed to form by cooling and
condensation of the baryons within haloes (White \& Rees 1978).
Thus understanding evolution of the abundance and clustering of 
dark haloes is an important first step towards understanding how
visible populations of objects form and cluster.
Because the growth of haloes is
purely gravitational, it is a relatively simple process.
Accurate analytic formulae are now available for many
properties of the halo distribution.
In this paper, we assemble these formulae and use them to
produce plots which give considerable insight into the growth
of nonlinear structure in the standard $\Lambda$CDM model. 
While none of the formulae plotted are original to this work,
we believe our diagrams are useful because they 
highlight several under-appreciated 
properties of the standard paradigm which have substantial impact
on its predictions for high redshift evolution.

\section {The key formulae}

Following common practice, we define the characteristic properties 
of a dark halo within a sphere of radius $r_{200}$ chosen so that 
the mean enclosed density is 200 
times the mean cosmic value $\overline{\rho}$. (Note that
other authors often use $r_{200}$ to denote the radius within which
the mean density is 200 times the {\it critical} value.)
With this definition, the mass and circular velocity of the halo
are related to $r_{200}$ by
\beq\label{r200}
r_{200}=\left[{GM\over 100 \Omega_{\rm m}(z) H^2(z)}\right]^{1/3},
~~\mbox{and}~~~
V_c=\left({GM\over r_{200}}\right)^{1/2},
\eeq
where $H(z)$ is Hubble's constant at redshift $z$, 
and $\Omega_{\rm m} (z)= \Omega_{\rm CDM}(z) + 
\Omega_{\rm B}(z)$ is the corresponding 
density parameter of non-relativistic matter. 
These quantities are related to their present-day values by
\beq
H(z)=H_0 E(z)\,
~~~\mbox{and}~~~~
\Omega_{\rm m}(z)={\omnow (1+z)^3\over E^2(z)}\,,
\eeq
where 
\beq
E(z)=\left[\ovnow+(1-\Omega_0)(1+z)^2
+\omnow (1+z)^3\right]^{1\over 2}.
\eeq
We define the characteristic or `virial' temperature of the halo to be 
\beq 
T={\mu V_c^2\over 2 k} = 3.6\times 10^5 \left[{V_c\over 100 {\rm
km~s^{-1}}}\right]^2 {\rm K}\,,
\eeq
where $\mu\approx 0.6 m_{\rm p}$ (with $m_{\rm p}$ being the
proton mass) is the mean molecular weight. Notice that even
in equilibrium the actual temperature of gas within the halo 
will differ from this virial temperature by a factor of order unity
which depends on the halo's detailed internal structure.
For a given mass we also use the current mean density of the
Universe $\overline{\rho}_0$ to define a radius 
\beq
R(M)\equiv 
\left({3M\over 4\pi {\overline\rho}_0}\right)^{1/3}\,,
\eeq  
which is the Lagrangian radius of the halo at the present time.  
Note that $M$ and $R$ are equivalent for a given cosmology.

In a Gaussian density field, the statistical properties
of dark matter haloes of mass $M$ depend on redshift and on 
\beq
\sigma^2 (R)
={1\over 2\pi^2}
\int_0^\infty 
k^3 P(k) {\tilde W}^2 (kR) {\rmd k\over k}\,, 
\eeq
where ${\tilde W}(x)= 3(\sin kR-kR\cos kR)/(kR)^3$ 
is the Fourier transform 
of a spherical top-hat filter with radius $R$, and $P(k)$ is 
the power spectrum of density fluctuations extrapolated to
$z=0$ according to linear theory. Assuming 
$\Omega_{\rm B, 0}\ll \Omega_{\rm CDM, 0}$, the CDM
power spectrum can be approximated by 
\beq
P(k)\propto k T^2 (k)\,,
\eeq
where we assume $n=1$ and $T(k)$, the transfer function representing
differential growth since early times, is 
\beq
T(k)=
{\ln (1+2.34 q)\over 2.34 q}
\left[1+3.89q+(16.1q)^2+(5.46q)^3+(6.71q)^4\right]^{-1/4}\,,
\eeq
with $q=k/[(\Omega_{\rm CDM, 0}+\Omega_{\rm B, 0}) h^2 \mpc^{-1}]$
(Bardeen et al. 1986). 

According to the argument first given by Press \& Schechter 
(1974, hereafter PS), the abundance of haloes as a function of
mass and redshift, expressed as the number of haloes per
unit comoving volume at redshift $z$ with mass in the interval
$(M, M+dM)$, may be written as
\beq\label{PS_diff}
n(M,z)\rmd M
=\sqrt{2\over \pi}{{\overline\rho_0}\over M}
{\rmd\nu\over\rmd M}\exp\left(-{\nu^2\over 2}\right)\,\rmd M\, .
\eeq
Here $\nu\equiv \delta_c/[D(z)\sigma (M)]$, where
$\delta_c\approx 1.69$ is a constant 
(we adopt $\delta_c=1.69$ throughout our discussion),
and the growth factor for linear
fluctuations can, following Carroll, Press \& Turner (1992) be 
taken as  $D(z)=g(z)/[g(0)(1+z)]$ with
\beq
g(z)\approx 
{5\over 2}\Omega_{\rm m}\left[\Omega_{\rm m}^{4/7}-\Omega_\Lambda
+(1+\Omega_{\rm m}/2)(1+\Omega_\Lambda/70)\right]^{-1}\,,
\eeq
and
\beq
\Omega_{\rm m}\equiv\Omega_{\rm m} (z)\,,
~~~
\Omega_\Lambda\equiv\Omega_\Lambda (z)=
{\ovnow\over E^2(z)}\,.
\eeq

Press \& Schechter derived the above mass function from the {\it
Ansatz} that the fraction $F$ of all cosmic mass which at redshift
$z$ is in haloes with masses
exceeding $M$ is {\it twice} the fraction of randomly placed spheres
of radius $R(M)$ which have linear overdensity at that time
exceeding $\delta_c$, the value at which a spherical perturbation
collapses. Since the linear fluctuation distribution is gaussian
this hypothesis implies
\beq\label{PS_F}
F(>M, z) ={\rm erfc}\left({\nu\over \sqrt{2}}\right)\,,
\eeq
and equation (\ref{PS_diff}) then follows by differentiation.
Let us define a characteristic halo mass at each redshift,
$M_\star(z)$, by $\nu=1$ [i.e. by $\sigma (M_\star)=\delta_c/D(z)$].
Such haloes may be called $1\sigma$ haloes. In general,
haloes with $\nu=N$ may be called ${\rm N}\sigma$ haloes.
According to equation (\ref{PS_F}), the mass fractions 
in haloes more massive than the $1\sigma$, $2\sigma$ and 
$3\sigma$ levels are
\beq
F_{1\sigma}\approx 0.32\,,~~~~~     
F_{2\sigma}\approx 0.046\,,~~~~~    
F_{3\sigma}\approx 0.0027\,.    
\eeq   

Numerical simulations show that although the scaling properties 
implied by the PS argument hold remarkably well for a wide variety of
hierarchical cosmogonies, substantially better fits to simulated
mass functions are obtained if the error function in equation
(\ref{PS_F}) is replaced by a function of slightly different shape.
Sheth \& Tormen (1999) suggested the following modification of
equation (\ref{PS_diff})
\beq\label{n_SMT}
n(M,z)\rmd M
=A\left(1+{1\over \nu'^{2q}}\right)
\sqrt{2\over \pi}{{\overline\rho}\over M}
{\rmd\nu'\over\rmd M}\exp\left(-{\nu'^2\over 2}\right)
\,\rmd M,
\eeq
where $\nu'=\sqrt{a}\nu$, $a=0.707$, $A\approx 0.322$ and
$q=0.3$. [See Sheth, Mo \& Tormen (2001) and Sheth \& Tormen 
(2002) for a justification of this formula in terms of an
ellipsoidal model for perturbation collapse.] 
The fraction of all matter in haloes with 
mass exceeding $M$ can be obtained by integrating equation 
(\ref{n_SMT}). To good approximation,
\beq\label{SMT_F}
F(>M, z)\approx 0.4\left(1+ {0.4\over \nu^{0.4}}\right)
{\rm erfc}\left({0.85\nu\over\sqrt{2}}\right)\,
\eeq
in the range $0.1<\nu<10$. In this case, 
\beq
F_{1\sigma}\approx 0.22\,,~~~~~     
F_{2\sigma}\approx 0.047\,,~~~~~   
F_{3\sigma}\approx 0.0055\,.   
\eeq 
In a detailed comparison with a wide range of simulations, Jenkins
et al. (2001) confirmed that this model is indeed a good fit 
provided haloes are defined at the same density contrast {\it relative
to the mean} in all cosmologies. This is the reason behind our
choice of definition for $r_{200}$ in equation (\ref{r200}).

Starting from an extension of the PS formalism due to  
Bond et al. (1991), Mo \& White (1996) developed an analytic
model for the spatial clustering of dark matter haloes which they
tested extensively against large N-body simulations. 
They proved that at large separations the cross-correlation between
haloes and mass is simply a constant $b$ times the autocorrelation 
of the mass. At any given redshift, this bias factor for
haloes of mass $M$ can be written as
\beq\label{b_MW}
b(M,z)=1+{{\nu^2(M,z)-1}\over \delta_c}.
\eeq
[Cole \& Kaiser (1989) had derived this asymptotic formula earlier
from a different but closely related argument based on 
the ``peak-background split''.]
At large separation the halo-halo autocorrelation and the halo-mass
cross-correlation are then just:
$\xi_{\rm hh}(r, z)=b^2(M,z)\xi_{\rm mm}(r,z)$,
$\xi_{\rm hm}(r,z)=b(M,z) \xi_{\rm mm}(r,z)$, where $\xi_{\rm mm}(r,z)$ is the
autocorrelation of the mass at redshift $z$. For this same population
of redshift $z$ haloes, 
a second bias factor defined in Mo \& White (1996) is
\beq\label{b0_MW}
b_0(M,z)=1+  {D(z)\over\delta_c}\left[\nu^2(M,z)-1\right]\,.
\eeq
This relates the autocorrelation 
of the $z=0$ descendants of the haloes 
(and their cross-correlation with mass) to the 
mass autocorrelation at $z=0$:
$\xi_{\rm dd}(r,0)=b_0^2(M,z)\xi_{\rm mm}(r,0)$,
$\xi_{\rm dm}(r,0)=b_0(M,z)  \xi_{\rm mm}(r,0)$. 

As first shown by
Jing (1998), the original Mo \& White formulae suffer from similar
inaccuracies to the original PS mass function, and indeed the two
discrepancies are closely related. More precise formulae can be
obtained from the ellipsoidal collapse model:
\beq\label{b_SMT}
b=1+{1\over\delta_c}
\left[\nu'^2+ b\nu'^{2(1-c)}
-{\nu'^{2c}/\sqrt{a}\over \nu'^{2c}+b(1-c)(1-c/2)}\right]\,, 
\eeq
\beq\label{b0_SMT}
b_0=1+{D(z)\over\delta_c}
\left[\nu'^2+ b\nu'^{2(1-c)}
-{\nu'^{2c}/\sqrt{a}\over \nu'^{2c}+b(1-c)(1-c/2)}\right]\,, 
\eeq
where $\nu'=\sqrt{a}\nu$,
$a=0.707$, $b=0.5$ and $c=0.6$ (Sheth, Mo \& Tormen 2001). 
Numerical simulations show that both of
these revisions are substantially more accurate than their 
spherical counterparts, especially  for haloes with $M< M_\star$
(Jing 1998; Sheth \& Tormen 1999; Casas-Miranda et al. 2002).
In the next section, we will use the
{\it rms} fluctuations in spheres of comoving radius $8\mpch$ as a measure
of the strength of halo clustering. Thus we define
\beq\label{Del_SMT}
\Delta_8(M,z) \equiv  \overline{b}(M,z) \sigma_8 D(z)
~~\mbox{and}~~~ 
\Delta_{8,0}(M,z)\equiv \overline{b}_0(M,z)\sigma_8\,,
\eeq 
to represent the clustering strength of haloes more massive than
$M$ at redshift $z$ and that of their $z=0$ descendants respectively.
The overbars on $b$ and $b_0$ in these two formulae represent
the fact that the values given by equations (\ref{b_SMT}) and
(\ref{b0_SMT}) are averaged over the distribution of haloes more
massive than $M$ at redshift $z$ [i.e. using equation (\ref{n_SMT})].
 
\section {The clustering pattern}

\begin{figure*}
\centering  
\vskip-1.0cm
\psfig{figure=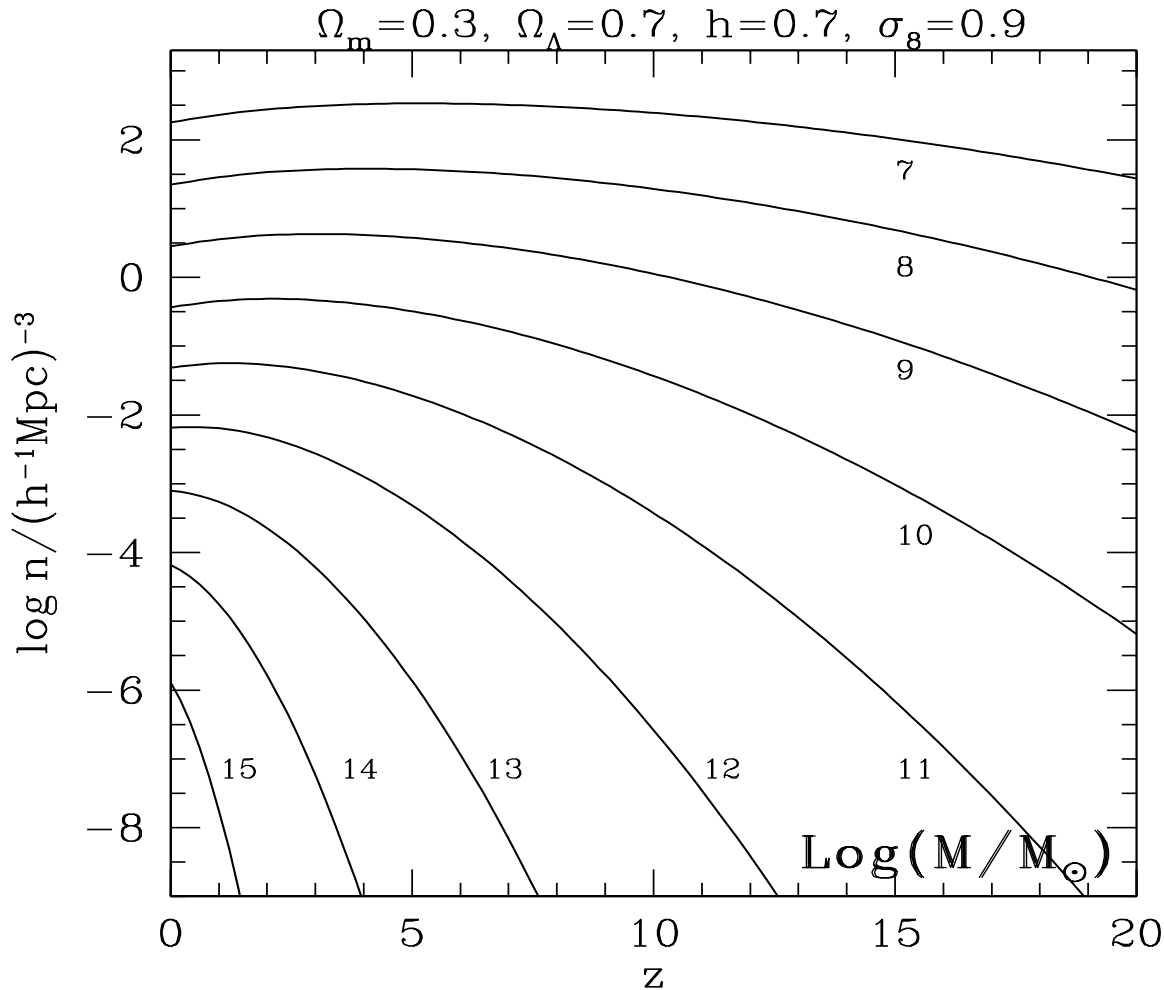,width=16.cm,height=16.cm,angle=0}
\vskip-1.0cm
\caption{Each curve indicates the variation with redshift of the
comoving number density of dark matter haloes with masses exceeding 
a specific value $M$ in the standard $\LCDM$ model with $\omnow=0.3$,
$\ovnow=0.7$, $h=0.7$ and $\sigma_8=0.9$. The label on each
curve indicates the corresponding value of $\log (M/\msun)$.}
\label{figure1}
\end{figure*}
\begin{figure*}
\centering  
\vskip-0.5cm 
\psfig{figure=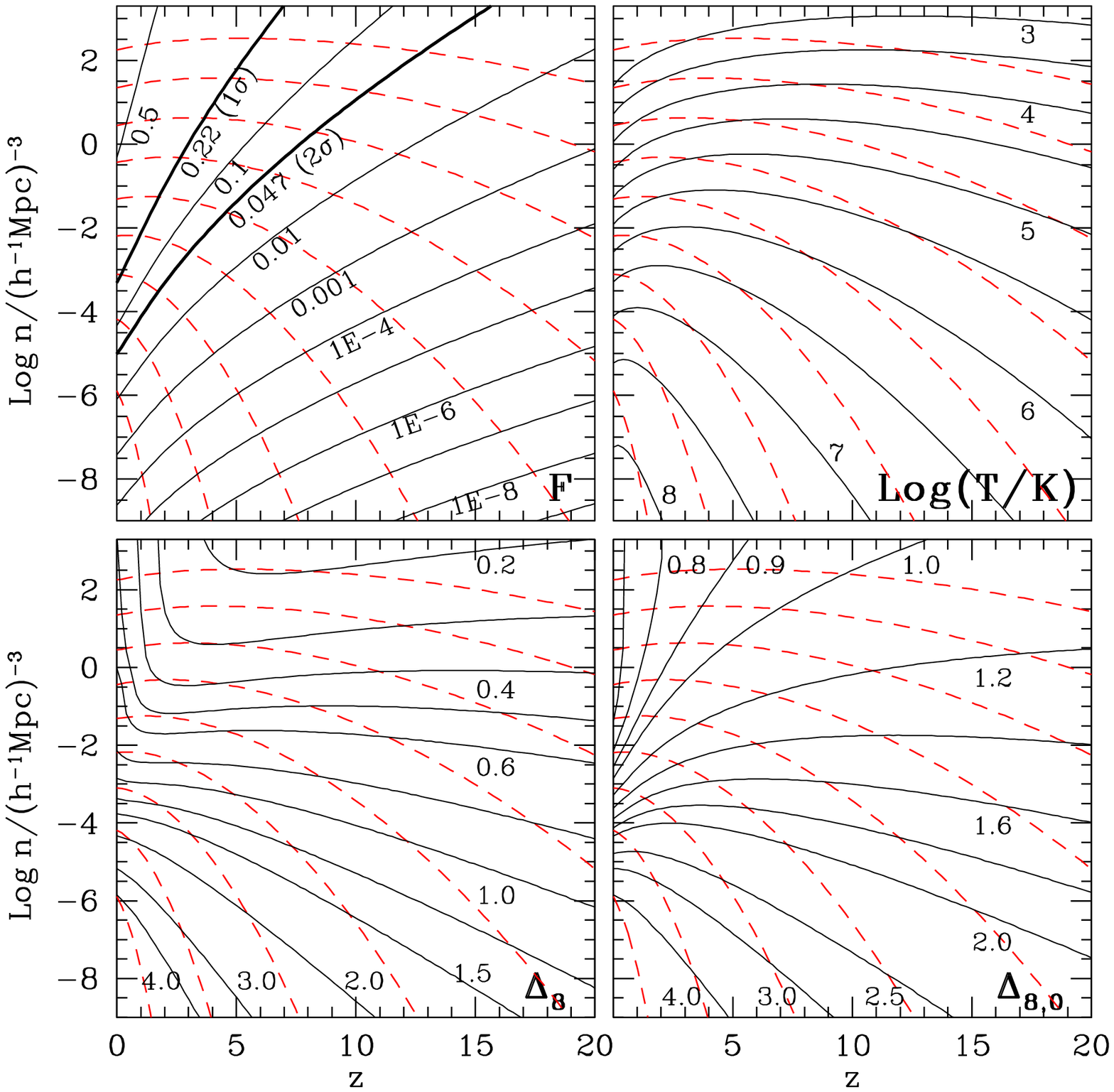,width=16.cm,height=16.cm,angle=0}
\vskip-1.0cm
\caption{The comoving number density of dark matter haloes with mass
exceeding $M(z)$ is plotted as a function of redshift 
for the standard $\LCDM$ model, $\omnow=0.3$,
$\ovnow=0.7$, $h=0.7$ and $\sigma_8=0.9$, and for various definitions of
the limiting mass $M(z)$. In the panel at top left, solid lines show
the halo abundance when $M(z)$ is chosen to define haloes
containing a given fraction $F$ of all cosmic matter; the values of $F$
are given as labels against each curve. The particular values of $F$
corresponding to ``$1\sigma$'' (or $M_*$)  and ``$2\sigma$'' haloes
are noted. The panel at top right shows results when $M(z)$ is chosen
to correspond to a given virial temperature at each redshift; labels
correspond to the logarithm of the temperature in Kelvin.
For the panel at bottom left, the limiting mass $M(z)$ is chosen so that the 
comoving clustering length of haloes (as measured by $\Delta_8$) has
a given value, shown by the label on each solid curve. Finally, for the
panel at bottom right, $M(z)$ is chosen so that the clustering strength
of the halo descendants at $z=0$ (as measured by $\Delta_{8, 0}$) has
a given value listed by the label on each curve. In all these plots,
dashed lines repeat the curves of Figure 1 and so can be used to
assign a limiting mass $M(z)$ to each
point in the abundance-redshift plane.
}
\label{figure2}
\end{figure*}
In this section we plot the halo abundances predicted by the analytic 
formulae given above for a version of the current ``concordance''
$\Lambda$CDM model. Specifically we pick a model with $\omnow=0.3$,
$\ovnow=0.7$, $h=0.7$ and $\sigma_8=0.9$. We have chosen to show
comoving abundance as a function of redshift for haloes with masses 
exceeding $M(z)$ for various definitions of the mass
limit $M(z)$. We have found these plots to be particularly
instructive, and by comparing them it is possible to read off a wide
variety of halo properties as a function of redshift. In the following
we will consider halo abundance as a function of halo mass, of total
halo contribution to the cosmic density, of halo virial temperature,
of halo clustering strength, and of the clustering
strength of the present-day descendants of high redshift haloes.

\subsection{Halo abundance as a function of mass}

In Figure 1 we show the comoving abundance of haloes of given mass
as a function of redshift for a wide range of halo masses. Plots of
this kind have appeared before in the literature a number of times
(e.g. Efstathiou \& Rees 1988; Cole \& Kaiser 1989). 
Our figure updates earlier versions
in that it is appropriate for the currently popular model and it
uses more accurate formulae (equation \ref{n_SMT}) than the 
Press-Schechter model used by previous authors. It is also required 
for comparison with the other plots we discuss below.

Figure 1 illustrates a number of well known properties of the
standard $\Lambda$CDM model. Haloes as massive as a rich galaxy
cluster like Coma ($M\sim 10^{15}\msun$) have an average
spacing of about $100h^{-1}$Mpc today, but their abundance drops 
dramatically in the relatively recent past. By $z=1.5$ it is already
down by a factor exceeding 1000, corresponding to a handful of objects 
in the observable Universe. The decline in the abundance of haloes
with mass similar to that of the Milky Way ($M\sim 10^{12}\msun$)
is much more gentle. By $z=5$ the drop is only about one order of
magnitude. At the smallest masses shown ($M\sim 10^7$ to $10^8\msun
$) there is little change in abundance over the full redshift
range $0<z<20$ that we plot. Notice also that the abundance of such
low mass haloes is actually declining slowly at low redshifts as 
members of these populations merge into larger systems faster than new
members are formed. It is interesting that haloes of mass $10^9\msun$
are as abundant at $z=20$ as $L_*$ galaxies are today, and haloes of
$10^{10}\msun$ are as abundant as present-day rich galaxy clusters.
Thus a significant population of relatively massive objects could, in
principle, be present even at these early times.

\subsection{The cosmic mass fraction in massive haloes}

For the upper left plot of Figure 2 we have used equation
(\ref{SMT_F}) to compute lower mass limits $M(z)$ such that the 
halo population contains a given fixed fraction of all cosmic mass.
As noted in Section 2, for our ``improved'' Press-Schechter models
haloes corresponding to $>1\sigma$ and $>2\sigma$ initial fluctuations
contain 0.22 and 0.047 of the cosmic mass respectively, independent of
redshift. By our conventions, the $1\sigma$ limiting mass is 
defined as $M_*$, the characteristic mass of clustering. By comparing
the $1\sigma$ line in Figure 2 with the constant mass lines copied
over from Figure 1, one can see that $M_*$ drops from just over
$10^{13}\msun$ at the present day to just below $10^7\msun$ at
$z=6$.

  From the redshift zero axis on this same plot, we can see that one percent
of all mass today is in objects more massive than $10^{15}\msun$, ten
percent is in objects more massive than $10^{14}\msun$, but more than
half is in objects {\it less} massive than $10^{10}\msun$, one
percent of the halo mass of the Milky Way. According to our standard
model, the mass of the Universe is currently distributed over objects
with an extremely wide range of masses. The predicted distribution of
the lower half of the mass is actually very uncertain, since use of our
formulae in this regime would involve extrapolation far below the
limits for which they have been tested against $N$-body simulations
(see Jenkins et al. 2001). For this reason we do not give curves for
mass fractions above 0.5.

At redshift 5 just under one percent of all matter is in haloes more
massive than that of the Milky Way, but by redshift 10 this fraction
has dropped to $10^{-6}$. At redshift 10 there is, nevertheless, still
one percent of all matter in haloes more massive than about
$10^{10}\msun$ and five percent in haloes more massive than about
$10^8\msun$. By redshift 20 only about $10^{-3}$ of all matter is
in haloes more massive than $10^8\msun$ and only about $10^{-6}$ in
haloes more massive than $10^{10}\msun$.

\subsection{Halo abundance as a function of temperature}

Because we define the boundary of our haloes at fixed
overdensity relative to the cosmic mean, the mass we assign to
a given dark halo will vary with redshift even in the absence of
evolution. This is actually quite a strong effect. For example,
we would assign an order of magnitude smaller mass to a halo
like the Milky Way's at redshift 5 than at the present day, even if
its density profile was unchanged. As a result, and also
because both the ionisation and the cooling of diffuse gas
within a halo depend primarily on its temperature, it is often more
helpful to study the abundance of haloes as a function of 
characteristic temperature (or equivalently of characteristic
circular velocity) rather than of mass. We give the relationship
between these quantities in equations (1) and (4) above, and we 
compare the evolution of halo abundance at fixed temperature to
that at fixed mass in the upper right panel of Figure 2. Again, we are
not the first to make such abundance evolution plots,
but the current plots update earlier versions by using
more accurate formulae and the current ``best bet'' cosmological
parameters.

The first point to note from this diagram is that made above -- haloes
of a given temperature correspond to different masses at different
redshifts. Thus a $10^6$K halo (with $V_c\sim 200$km/s) has a mass of 
about $2\times 10^{12}\msun$ at the present day, but a mass of only
$2\times 10^{10}\msun$ at $z=20$. Conversely, a $2\times 10^{10}\msun$ 
halo at $z=0$ has a temperature of only $5\times 10^4$K and so a circular
velocity of only 36 km/s.

A second important point is that over a wide range of temperature
the evolution of abundance with redshift is quite slow. Thus the 
abundance of haloes hotter than 1keV is approximately the same at 
$z=3$ as at $z=0$. Indeed, for temperature limits equal to or below 
the observed temperatures of Abell clusters ($T\la 5$keV) the 
abundance of objects actually increases with increasing redshift 
away from $z=0$. An X-ray temperature-selected sample of galaxy
clusters is expected to show little abundance evolution 
unless the temperature threshold for inclusion is quite high.
For the characteristic temperature of the Milky Way's halo, $T\sim
10^6$K corresponding to $V_c\sim 200$km/s, the abundance
of systems at $z=8$ is about the same as it is at $z=0$. Apparently
the formation of galaxies like our own could, in principle, start at
very early times.

Infall onto haloes with characteristic temperatures greater than 
$10^4$K will produce strong enough shocks to ionise the
infalling gas. As a result, atomic line cooling is expected to be
efficient in such systems and to lead to condensation of dense gas
at their centres, possibly with associated star formation. Figure 2
shows the abundance of such systems to be almost constant at
$n\sim 10h^3$Mpc$^{-3}$ all the way from $z=20$ down to $z=3$. At
$z=15$ about 1\% of all matter is already in objects above the
collisional ionization threshold, and by $z=5$ this fraction has
climbed to about 20\%. In the absence of effects other than cooling,
roughly  1 and 20\% of all baryons would be in dense systems by redshifts of 
15 and 5 respectively, whereas current estimates suggest that this
fraction is below 10\% even at $z=0$ (Balogh et al. 2001). In fact, it
has long been argued that radiative and hydrodynamic feedback must be
associated with the formation of stars and active galactic nuclei, and
that it must limit the condensation of gas within smaller haloes (White \&
Rees 1978; White \& Frenk 1991; Efstathiou 1992).

According to the calculations of Gnedin (2000) for standard 
reionisation models, haloes with $T\la 10^5\Kdegree$ are unable to trap, 
and therefore to cool,
significant amounts of diffuse photoionised gas at low redshifts ($z\la
5$). Since reionisation is known to have occurred before $z=6$, Figure
2 suggests that more than half of all baryons 
were never part of a halo in which cooling was efficient. These baryons 
must currently reside in a diffuse intergalactic medium (see Cen
\& Ostriker 1999).

\subsection{Halo clustering}

As noted in Section 2 we have decided to characterise the clustering
of haloes as a function of mass and redshift using $\Delta_8(M,z)$
the {\it rms} overdensity in the number of haloes more
massive than $M$ at redshift $z$ after smoothing with a spherical
top-hat filter of comoving radius $8h^{-1}$Mpc. This measure is
convenient to calculate and has become traditional because it is 
close to unity for moderately bright galaxies in the present Universe. The
theory of Mo \& White (1996), as corrected empirically by Jing (1998)
and Sheth \& Tormen (1999), shows that at each redshift and for
halo masses below $M_*$, the value of $\Delta_8(M,z)$ varies little 
with $M$ and is just below $D(z)\sigma_8$, the corresponding mass density
fluctuation. For halo masses above $M_*(z)$ the value of $\Delta_8$
increases rapidly with $M$ [see equations (\ref{b_SMT}) and
(\ref{Del_SMT})].

The lower left panel of Figure 2 shows the abundance-redshift relation
for haloes more massive than $M(z)$ chosen so that 
$\Delta_8$ takes specific values, given as labels beside 
each curve. From the $y$-axis of this plot we see that no halo
population in the present Universe has $\Delta_8 < 0.5$, that 
haloes more massive than $M_*$ currently 
have $\Delta_8\sim 1.2$, and that rich galaxy
clusters at $z=0$ have $\Delta_8\ga 3$. The locus of $M_*$ haloes
is clearly visible in this plot joining the points of maximum 
curvature on each of the constant $\Delta_8$ curves. (Compare with the
$1\sigma$ curve in the upper left panel of Figure 2.)

A striking feature of this clustering plot is the weak dependence
of clustering strength on redshift for haloes more massive than
$M_*(z)$. In this regime the increase in bias with increasing redshift
compensates for the decreasing clustering strength of the underlying
mass distribution. Thus haloes with the abundance of $L_*$ galaxies,
$n\sim 10^{-3}h^3{\rm Mpc}^{-3}$, have $\Delta_8\sim 0.9$ for all
$z<10$. Over this redshift range their mass drops from about
$10^{13}\msun$ at $z=0$ to about $10^{11}\msun$ at $z=10$. It is 
interesting that their clustering
strength is close to that measured at $z\sim 3$ for the ``Lyman
break'' galaxy population identified by Steidel et al 
(1996; see also Adelberger et al. 1998) which 
indeed has a comoving number density of about $0.001h^3{\rm Mpc}^{-3}$.
This has led many authors, beginning with Mo \& Fukugita
(1996), to speculate that Lyman break galaxies may be the central objects of
$z\sim 3$ haloes and that their UV luminosity may correlate well with
their halo mass. The halo mass of a Lyman break galaxy at $z\sim 3$
can then be read off from Figure 2; it is $M\sim10^{12}\msun$. Clearly
a more detailed study of the clustering of high redshift galaxies
should provide detailed information about how the masses of the haloes
they inhabit are related to their observable properties (Baugh et al.
1998; Mo, Mao \& White 1999; Somerville, Primack \& Faber 2001;
Wechsler et al 2001; Shu, Mao \& Mo 2001). 

Another surprising property of the clustering predictions is that
haloes of given mass actually become {\it more
strongly} clustered with increasing redshift once the mass chosen
exceeds $M_*(z)$. The same is also true, although more weakly so,
for samples selected to a fixed virial temperature.
Thus galaxy clusters selected to a fixed rest-frame X-ray 
temperature limit are expected to be equally clustered at $z=1$ 
and at $z=0$, while clusters selected to a given observed temperature
limit will be substantially more clustered at the higher 
redshift. Relatively low mass haloes can be surprisingly strongly 
clustered even at very high redshift. For example, at $z=20$ haloes
of mass above $10^8\msun$ have a virial temperature above 
$2\times 10^4$K, contain 0.1\% of all matter, 
and have a clustering strength of $\Delta_8\sim 0.4$.

\subsection{Clustering of the present-day remnants of high redshift
haloes}

One of the major puzzles in current astrophysics is the relation
between the objects observed in the high redshift Universe and those
around us today. Which present-day galaxies contain the stellar
population we see forming in Lyman break galaxies? Which host the massive
black holes which powered high redshift quasars? Where are the
remnants of the very first stellar populations? How are the heavy
elements they produced distributed through the present Universe?
Our current understanding of where to look for such remnants 
comes primarily from simulations (Governato et al 1998; Cen \& 
Ostriker 1999; White \& Springel 2000; Kauffmann \& Haehnelt 2001) 
but considerable intuition can also be gained from simpler analytic
estimates of remnant clustering (Mo \& Fukugita 1996). We illustrate
this here using $\Delta_{8,0}(M,z)$, the clustering strength of the
present-day descendants of halos with mass greater than $M$ at
redshift $z$. Note that this measure weights each descendant by the
number of its high redshift progenitors.

In the lower right panel of Figure 2 we give abundance-redshift
relations for haloes selected to mass limits $M(z)$ which give 
the specific values of $\Delta_{8,0}$ listed against each curve.
Along the $z=0$ axis the abundance-clustering strength relation in
this plot is, of course, identical to that of the $\Delta_8$ plot
in the lower left panel. At higher redshifts $\Delta_{8,0}$ exceeds
$\Delta_8$ everywhere as a result of the growth of clustering with
time. The $\Delta_{8,0}=0.9$ curve is close to the 
curve for $F=0.22$ in the top left panel; at each redshift $1\sigma$ 
haloes are unbiased relative to the mass, and their descendants
remain unbiased as clustering evolves. Over most of our abundance-redshift
plot, $\Delta_{8,0}$ is substantially larger than 0.9, so that the
the remnants of the relevant halo populations are more clustered today
than the dark matter or than $L_*$ galaxies. The descendants
of $z=3$ halos with the abundance of Lyman break galaxies 
($n\sim 10^{-3}h^3{\rm Mpc}^{-3}$) have $\Delta_{8,0}\sim 1.6$,
suggesting that the Lyman break systems have evolved preferentially
into massive early-type galaxies (Mo \& Fukugita 1996; Governato et
al. 1998).
 
Bright quasars have now been seen to redshifts beyond 6. 
Their observed comoving number density
is $n\sim 4\times 10^{-8} h^3{\rm Mpc}^{-3}$ at $z\sim 3$, 
and $n\sim 2\times 10^{-9} h^3{\rm Mpc}^{-3}$ at $z\sim 6$
(Fan et al. 2001). If we adopt a typical quasar lifetime 
at redshift $z$ of $3\times 10^{7} [H_0/H(z)]$ years
(e.g. Kauffmann \& Haehnelt 2000), 
the implied number density of host haloes is
$n\sim 10^{-5}h^3{\rm Mpc}^{-3}$ at $z\sim 3$,  
and $n\sim 10^{-6.5}h^3{\rm Mpc}^{-3}$ at $z\sim 6$. From Figure 1 
we see that the masses of such haloes 
are $M\sim 10^{13}\msun$, suggesting there will be no problem
forming such luminous objects by the observed redshifts in our
standard cosmology.  At $z\sim 3$ these haloes have 
correlation strength $\Delta_8\sim 1.5$
corresponding to a comoving correlation length of about $8\mpch$
\footnote{We have approximated the correlation function 
as $\xi(r)\sim \Delta_8^2  (r_0/r)^{1.8}$, 
so that the correlation length is $r_0 \Delta_8^{2/1.8}$,
where $r_0\sim 5\mpch$ is the correlation length for the mass.}.
Their present-day descendants then have $\Delta_{8,0}\sim 2$, or a 
correlation length about $11\mpch$. The predicted correlation lengths
for the $z\sim 6$ quasars and their descendants are about $11\mpch$ 
and $14\mpch$ respectively. These $z=0$ correlation lengths are 
comparable to those of the most luminous elliptical galaxies in the
present Universe. Such galaxies are indeed now thought to host
the $\sim 10^9\msun$ black holes which must have powered these
distant quasars (Gebhardt et al 2000; Ferrarese \& Merritt 2000).
Notice that our inferred clustering lengths are quite sensitive to 
the quasar lifetimes we assumed. A number of authors have pointed
out that clustering can therefore constrain quasar 
lifetimes (La Franca, Andreani \& Cristiani 1998; Fang \& Jing 1998;
Haehnelt, Natarajan \& Rees 1998;
Haiman \& Hui 2001; Martini \& Weinberg 2001; 
Kauffmann \& Haehnelt 2001).

High-$z$ quasars are also expected to be strongly correlated with
other objects at the same redshift. For example, at 
$z\sim 6$ the mean density enhancement of haloes with 
$M\sim 10^{10}\msun$ in an $8\mpch$ sphere surrounding a bright
quasar is the product of the $\Delta_8$ values for the two populations,
i.e. about $0.4 \times 2 = 0.8$. Such dense environments surrounding
quasars may have important implications for the 
interpretation of quasar absorption spectra, in particular for
the proximity effect or for the effective
absorption optical depth of the foreground IGM.

As a final example of these clustering plots, we consider the 
distribution of the metals
produced by the first generations of stars. Tegmark et al. (1997, 
see also Loeb \& Barkana 2001 and references therein)
showed that primordial gas in haloes with virial temperatures 
above $3\times 10^3\Kdegree$ can cool by molecular hydrogen and atomic
line cooling at $z\la 20$. Population III stars may form in such 
haloes, which would then begin ionising the intergalactic medium and
polluting it with heavy elements. As one can see from
Figure 2, just below 1 percent of the cosmic mass is already 
in such haloes at $z\sim 20$, and their clustering is already
moderately strong, $\Delta_8\sim 0.25$. By redshift zero the metals
produced by this population will have
$\Delta_{8,0}\sim 1.15$ and so will be as clustered as $L_*$ galaxies.
By redshift 3 the clustering strength of these metals has
already reached $\Delta_8\sim 0.6$. Thus the metallicity
distribution in the gas probed by observed Lyman $\alpha$ forest
is predicted to be highly inhomogeneous even if enrichment is due to
Population III objects burning at early times. [These various clustering
strengths can be read off Figure 2 as follows. The upper left panel
gives $n\sim 10^2h^3{\rm Mpc}^{-3}$ as the abundance of objects at
$z=20$ with a virial temperature above $3\times 10^3$K. The panels at lower
left and lower right then give the clustering strength of these
objects and of their $z=0$ descendants respectively. The clustering 
strength of the $z=3$ descendants is obtained by following
the $\Delta_{8,0}=1.15$ line in the bottom left panel from $z=20$ down
to $z=3$ where it corresponds to $n\sim 10^{-1.5}h^3{\rm Mpc}^{-3}$.
The appropriate clustering strength $\Delta_8\sim 0.6$ is then read
off from the bottom left panel.]

\section{Concluding words}

Throughout this paper we have discussed a single version of the
current standard $\Lambda$CDM model, but the formulae given
in Section 2 are complete enough that the reader can easily produce
modified versions of our plots for other choices of the cosmological
parameters. These can be useful for understanding whether the abundance 
and clustering of a population of interest may be sensitive to the
particular cosmological context in which it is evolving. We have given
examples of applications to galaxy clusters, to Lyman break galaxies,
to quasars, and to the enrichment of the intergalactic gas by
high redshift Population III stars. Other applications are 
possible, but these already illustrate how the observed abundance and
clustering of a population can be used to infer the masses of the dark
haloes in which it is embedded. Even for the ``standard'' model we
have studied, a number of the properties evident from our plots
seem at first sight to be counter-intuitive or surprising. We have
found this graphical presentation of the model properties to
be surprisingly useful, and we hope that others will also.


{}

\end{document}